\documentclass[prl,twocolumn,amsmath,amssymb,floatfix,reprint,footinbib,superscriptaddress,longbibliography,showkeys]{revtex4-1}
\usepackage{dcolumn}
\usepackage{graphicx}
\usepackage{mathrsfs}
\usepackage{mdwlist}
\usepackage{subfigure}
\usepackage{booktabs}
\usepackage{multirow}
\usepackage{amsmath}
\usepackage{textcomp}
\usepackage{upgreek}
\usepackage{dsfont}
\usepackage{amstext}
\usepackage{amssymb}
\usepackage{amsbsy}
\usepackage{appendix}
\usepackage{soul}
\usepackage{threeparttable}
\usepackage{bbm}
\usepackage{amsthm}
\usepackage{graphicx}
\usepackage{lipsum}
\usepackage{xcolor}
\usepackage[none]{hyphenat}
\usepackage{textcomp}
\usepackage{color}
\usepackage[colorlinks,citecolor=blue]{hyperref}
\definecolor{Dgreen}{RGB}{0, 100, 0}
\usepackage{url}
\usepackage[colorlinks]{hyperref}
\hypersetup{%
	plainpages=true,
	breaklinks=true,  %not default in dvips mode, so we must specify
	hypertexnames=false,  %not ideal, but needed when pagenums duplicate (`i' vs. `1')
	pageanchor=true,
	colorlinks=true,
	linkcolor={blue},
	citecolor={red},
	urlcolor={blue},
	anchorcolor={black}
}

\begin{document}
%\begin{CJK*}{UTF8}{gbsn}

\title{Error-Tolerant Amplification and Simulation of the Ultrastrong-Coupling Quantum Rabi Model}
\author{Ye-Hong Chen}
\affiliation{Fujian Key Laboratory of Quantum Information and Quantum Optics, Fuzhou University, Fuzhou 350116, China}
\affiliation{Department of Physics, Fuzhou University, Fuzhou 350116, China}
\affiliation{Theoretical Quantum Physics Laboratory, RIKEN Cluster for Pioneering Research, Wako-shi, Saitama 351-0198, Japan}

\author{Zhi-Cheng Shi}
\affiliation{Fujian Key Laboratory of Quantum Information and Quantum Optics, Fuzhou University, Fuzhou 350116, China}
\affiliation{Department of Physics, Fuzhou University, Fuzhou 350116, China}

\author{Franco Nori}%\thanks{fnori@riken.jp}%
\affiliation{Theoretical Quantum Physics Laboratory, RIKEN Cluster for Pioneering Research, Wako-shi, Saitama 351-0198, Japan}%
\affiliation{Quantum Information Physics Theory Research Team, RIKEN Center for Quantum Computing, Wako-shi, Saitama 351-0198, Japan}%
\affiliation{Department of Physics, University of Michigan, Ann Arbor, Michigan 48109-1040, USA}

\author{Yan Xia}\thanks{xia-208@163.com}
\affiliation{Fujian Key Laboratory of Quantum Information and Quantum Optics, Fuzhou University, Fuzhou 350116, China}
\affiliation{Department of Physics, Fuzhou University, Fuzhou 350116, China}

\date{\today}

\begin{abstract}
{Cat-state qubits formed by photonic cat states have a biased noise
channel, i.e., one type of error dominates over all the others.
We demonstrate that such biased-noise qubits are also promising for error-tolerant simulations
of the quantum Rabi model (and its varieties) by coupling a cat-state qubit to
an optical cavity. Using the cat-state qubit can effectively enhance the counter-rotating coupling, allowing us to explore several fascinating quantum phenomena relying on the
counter-rotating interaction. Moreover, another benefit from biased-noise cat qubits is that the two main 
error channels (frequency and amplitude mismatches) are both exponentially suppressed.}
Therefore, the simulation protocols are robust against parameter errors of the parametric drive which 
determines the projection subspace. 
We analyze three examples: (i) collapse and revivals of quantum states; (ii) hidden symmetry and tunneling dynamics; and (iii)
pair-cat-code computation.

\end{abstract}

%\pacs {03.67.}

\maketitle

\textit{Introduction.}---The quantum Rabi model (QRM) has been used to describe the dynamics of
a wide variety of physical setups \cite{Scully1997Book,Agarwal2012Book,Auffeves2013Book}. Generally, the QRM can be divided into different
coupling regimes \cite{FriskKockum2019,FornDiza2019}, where
the most interesting one is the ultrastrong coupling (USC) regime.
This is because the USC can open new perspectives
for efficiently simulating known effects and observing fundamental phenomena in quantum nonlinear optics \cite{Scully1997Book,Agarwal2012Book,Auffeves2013Book,FriskKockum2019,FornDiza2019,Kockum2017Pra}. 
These coupling regimes are established when the light-matter
interaction energy is comparable to the bare frequencies of the uncoupled systems.

Though the USC regime has been achieved in several systems \cite{FornDaz2016Np,Yoshihara2016NP,Yoshihara2017PRA,Bosman2017Njpqi}, it is still difficult to study
unexplored physics and observe its fascinating quantum phenomena at will
because the coupling regimes should be implemented in a fully tunable and
efficient manner \cite{FornDiza2019,FriskKockum2019}. In this respect, proposals in both theory \cite{Ballester2012PRX,Qin2018PRL,Leroux2018PRL,Sanchez2020Pra} and experiments \cite{Braumller2017NC,Lv2018PRX,Cai2021NC,Chen2021NC,Zheng2023PRL} for 
analog quantum simulation \cite{Georgescu2014RMP,Qin2024Arxiv} of the QRM were put forward \cite{FriskKockum2019,FornDiza2019}.
Researchers can therefore study USC-induced quantum phenomena, such as the asymmetry of
the vacuum Rabi splitting \cite{Ashhab2010Pra,Cao2011Njp,Leroux2018PRL,Lv2018PRX}, nonclassical photon statistics, and superradiance transition \cite{Cai2021NC,Chen2021NC,Zheng2023PRL}.
For simulating the QRM, additional control fields are usually applied to effectively enhance the ratio between 
the coupling strength and the bare frequencies in a specific rotating frame to reach the USC regime.
However, the simulation protocols \cite{Ballester2012PRX,Leroux2018PRL,Qin2018PRL,Braumller2017NC,Sanchez2020Pra,Lv2018PRX,Cai2021NC,Chen2021NC,Zheng2023PRL,Georgescu2014RMP,Qin2024Arxiv} are sensitive to deviations in these additional drives because projecting the system
onto a wrong rotating frame can result in a totally wrong effective Hamiltonian.
For instance, in the protocols \cite{Qin2018PRL,Leroux2018PRL,Chen2021,Burd2021Np} of squeezing-induced USC,
a small deviation in the squeezing strength results in a totally different effective Hamiltonian and breaks the desired dynamical predictions.
Similar problems exist in other simulation protocols \cite{Qin2024Arxiv}.
 
For realizing an error-tolerant simulation of the QRM in the USC regime, 
we propose to use a logic qubit, e.g., the cat-state qubit, instead of a physical qubit.
The cat-state qubit \cite{Ralph2003,Gilchrist2004,Mirrahimi2014,Mirrahimi2016,Girvin2023Spln} formed by photonic cat states was introduced for fault-tolerant quantum computing because
it is noise-biased and experiences only
bit-flip noise \cite{Cai2021,Ma2021,PuriPrx2019,Puri2020,Grimm2020,Chen2022Prappl}. 
It can be realized by parametrically driving a Kerr-nonlinear resonator (KNR) \cite{Puri2017,PuriPrx2019,Puri2020,Chen2022Prappl,Grimm2020,WangPrx2019}.
The odd and even cat states are two degenerate eigenstates of this parametrically-driven KNR.
The coupling between the KNR and a cavity can be 
linearly enhanced when we treat the KNR as a cat-state qubit,
allowing to reach the USC regime.
Because the cat-state qubit preserves the noise bias, 
our simulation protocol is also noise-biased and can exponentially suppress 
the errors caused by deviations in the parametric drive.
As examples, we show how this method can explore the following phenomena in the USC regime:
(i) collapse and revivals; (ii) hidden symmetry and tunneling dynamics; (iii)
pair-cat-code computation.

%For instance, the simulation methods described in 
%Refs.~\cite{Ballester2012PRX,Lv2018PRX} require two additional drives to be applied to a Jaynes-Cummings model.

%superpositions of coherent states.
%The cat-state qubits are attractive because coherent states
%are eigenstates of the photon annihilation operator and
%therefore single-photon loss induces simple, tractable errors \cite{Heeres2017,Ofek2016}.

%As a result, one can apply efficient error-protection into the physical layer while maintaining
%simplicity.

\begin{figure}
	\centering
	\scalebox{0.37}{\includegraphics{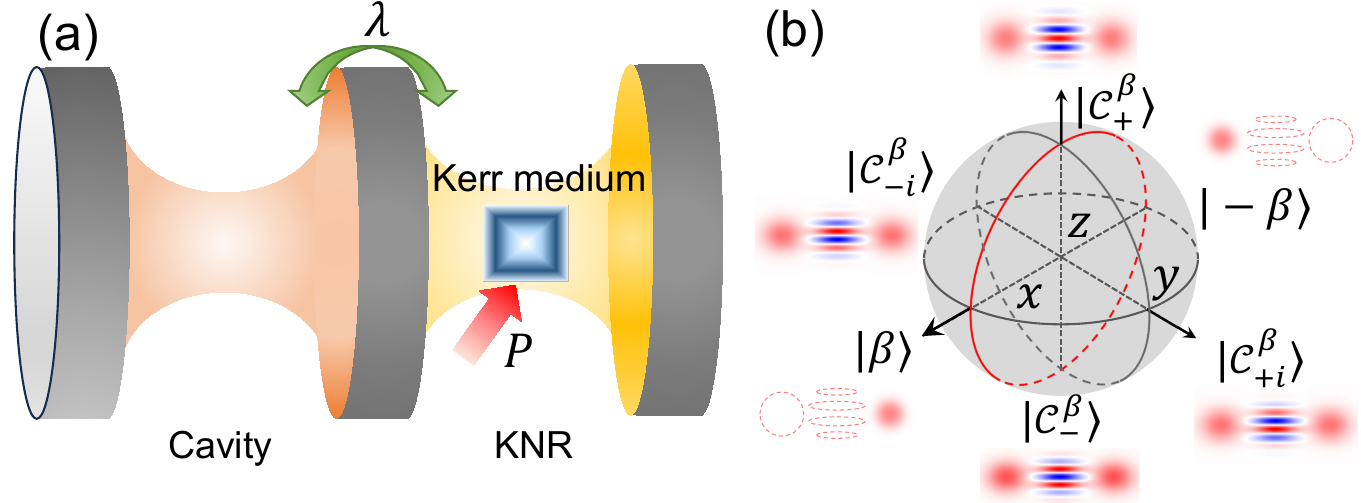}}
	\caption{{(a) Schematic of the setup: a parametrically driven
		Kerr-nonlinear resonator (KNR) is weakly
		coupled to a cavity with strength $\lambda$.} (b) Bloch sphere of the cat-state qubit for large $\beta$. The red circle denotes the only possible rotation direction for the qubit.
	}
	\label{fig0}
\end{figure}

\textit{Physical model.}---{As shown in Fig.~\ref{fig0}(a), we consider a general
physical model of a KNR of frequency $\omega_{\rm KNR}$ coupled to a cavity of frequency $\omega$, with coupling strength $\lambda$.
A two-photon drive (i.e., parametric drive) with amplitude $P$ and frequency $\omega_{p}$ is applied to the KNR.
Thus, working in a frame rotating at half the
parametric drive frequency, the Hamiltonian under the rotating-wave approximation becomes
\begin{align}\label{eq1}
	H=&H_{0}+H_{\rm KNR}+H_{\rm int},\cr
	H_{0}=&\Delta a^{\dag}a+\delta b^{\dag}b,\cr
    H_{\rm KNR}=&-Kb^{\dag 2}b^{2}+Pb^{\dag 2}+P^{*}b^{2},\cr
    H_{\rm int}=& \lambda ab^{\dag}+\lambda^{*}a^{\dag}b. 
\end{align}
Here, $\Delta=\omega-\omega_{p}/2$ and $\delta=\omega_{\rm KNR}-\omega_{p}/2$ are detunings, $K$ is the strength of the self-Kerr nonlinearity. A possible implementation of this Hamiltonian involves superconducting circuits (See the Supplemental Material \cite{SM} for details), which have experimentally realized Kerr-cat qubits \cite{WangPrx2019,Grimm2020} and showed a strong suppression of frequency
fluctuations due to $1/f$ noise for the pumped cat \cite{Grimm2020,Darmawan2021PrxQ,Xu2022book}.}

When the parameters $\Delta$, $\delta$, and $\lambda$ are far smaller than $K$ and $P$,
we can project the whole system onto the subspace spanned by the eigenstates of $H_{\rm KNR}$ \cite{Puri2017,PuriPrx2019,Chen2022Prappl}. Coincidentally, the ground eigenstate of the Hamiltonian $H_{\rm KNR}$
are a set of degenerate eigenstates
\begin{align}
	|\mathcal{C}_{\pm}^{\beta}\rangle=\frac{1}{\sqrt{\mathcal{N}_{\pm}}}\left(|\beta\rangle\pm|-\beta\rangle\right),
\end{align}
which are separated from the other eigenstates with an energy gap $E_{\rm gap}\simeq 4K|\beta|^{2}$ \cite{PuriPrx2019},
where $\beta=\sqrt{P/K}$ is the complex amplitude of the coherent state $|\beta\rangle$ 
and $\mathcal{N}_{\pm}$ are normalized coefficients.

\begin{figure}
	\centering
	\scalebox{0.32}{\includegraphics{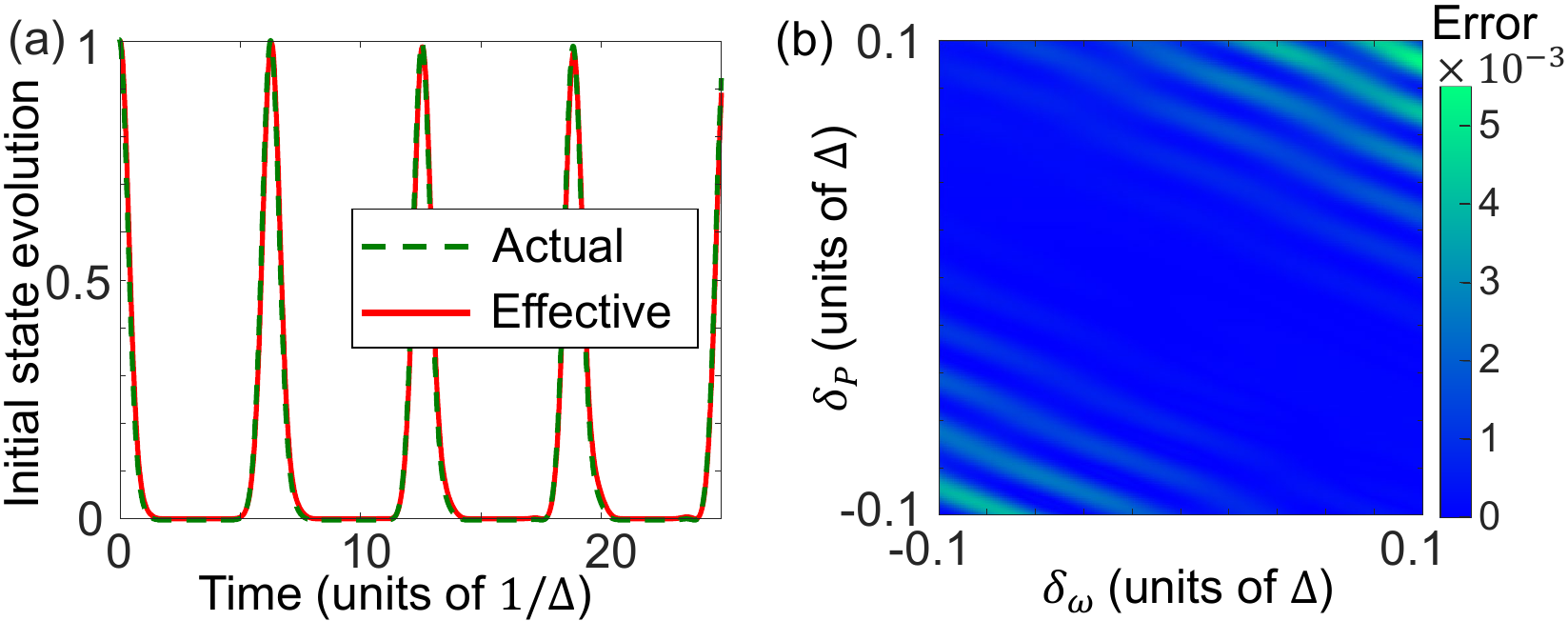}}
	\caption{(a) Time evolutions of the initial state $|0,\mathcal{C}_{+}^{\beta}\rangle$ governed by the effective Hamiltonian $H_{R}$ (red-solid curve) and the Hamiltonian $H$ (green-dashed curve). 
		(b) Deviations in the population of the initial state $|0,\mathcal{C}_{+}^{\beta}\rangle$ by adding the error Hamiltonian $H_{\rm err}$ after a finite-time evolution with $t=4\pi/\lambda$. We choose $\beta=2$, $K=10\Delta$, $\lambda=\Delta$, and $\delta= 0.1\Delta$ to reach the USC regime.
	}
	\label{fig1}
\end{figure}

In the limit of $\{\Delta,\delta,\lambda\}\ll E_{\rm gap}$,
if the KNR is initially in the cat-state subspace $\mathcal{C}=\{|\mathcal{C}_{\pm}^{\beta}\rangle\}$,
its dynamics will be confined to this subspace.
The KNR can be seen
as a two-level system, i.e., a cat-state qubit as shown in Fig.~\ref{fig0}(b).
We define the Pauli matrices $\sigma_{+}=|\mathcal{C}_{-}^{\beta}\rangle\langle\mathcal{C}_{+}^{\beta}|$,
$\sigma_{-}=\left(\sigma_{+}\right)^{\dag}$, and $\sigma_{z}=|\mathcal{C}_{-}^{\beta}\rangle\langle\mathcal{C}_{-}^{\beta}|-|\mathcal{C}_{+}^{\beta}\rangle\langle\mathcal{C}_{+}^{\beta}|$.
Working in the cat-state subspace, the effective Hamiltonian reduces to
\begin{align}\label{eq3}
	H_{R}=\Delta a^{\dag}a+\frac{\tilde{\delta}}{2}\sigma_{z}+\left[\lambda\beta\left( \frac{\sigma_{+}}{A}+A \sigma_{-}\right)a+{\rm H.c.}\right],
\end{align}
which describes a tunable anisotropic QRM. Here, $A=\sqrt{\tanh{|\beta|^2}}$  and 
 $\tilde{\delta}=2\delta|\beta|^{2}{\rm csch}(2|\beta|^2)$.
The unitary term $\mathbbm{1}_{\beta}=|\mathcal{C}_{-}^{\beta}\rangle\langle\mathcal{C}_{-}^{\beta}|+|\mathcal{C}_{+}^{\beta}\rangle\langle\mathcal{C}_{+}^{\beta}|$ is omitted in Eq.~(\ref{eq3}).
For large $\beta$, $H_{R}$ takes the standard form of the QRM with an enhanced coupling strength $g\simeq \lambda\beta$ 
because of $A\simeq A^{-1}\simeq 1$. 
As shown in Fig.~\ref{fig1}(a), the effective dynamics governed by the Hamiltonian $H$ coincides very well
with that governed by the effective Hamiltonian $H_{R}$.

\textit{Parameter errors.}---In our protocol,
there are mainly two errors:
(i) driving frequency mismatch described by $\delta_{\omega}b^{\dag}b$; and
(ii) driving amplitude imperfections, i.e., deviations in $P$ (or equivalently $K$).
Therefore, the error Hamiltonian is
\begin{align}
	H_{\rm err}=\delta_{\omega}b^{\dag}b+\delta_{P}b^{\dag 2}+\delta_{P}^{*}b^{2}.
\end{align}
Projecting onto $\mathcal{C}$, $H_{\rm err}$ becomes
\begin{align}\label{eq5}
	H_{\rm err}\approx \delta_{\omega}|\beta|^2\left(
	\begin{array}{cc}
		A^{-2} & 0\cr
		0 & A^{2}
	\end{array}
	\right)
	+\delta_{P}\left(\beta^2+\beta^{* 2}\right)\mathbbm{1}_{\beta},
\end{align}
which is approximatively a unit matrix for large $\beta$.
As long as $\delta_{\omega},\delta_{P}\ll E_{\rm gap}$, $H_{\rm err}$ only
causes a change in the global phase.
We demonstrate this in Fig.~\ref{fig1}(b) by illustrating 
the dynamics governed by the total Hamiltonian $\mathcal{H}=H+H_{\rm err}$. 
When the deviations reach
$\delta_{P}=\delta_{\omega}=\pm0.1\Delta$,
the deviation in the final state population is less than $0.5\%$.
%Therefore, our simulation protocol is insensitive to these dominant errors.

 \begin{figure}
 	\centering
 	\scalebox{0.39}{\includegraphics{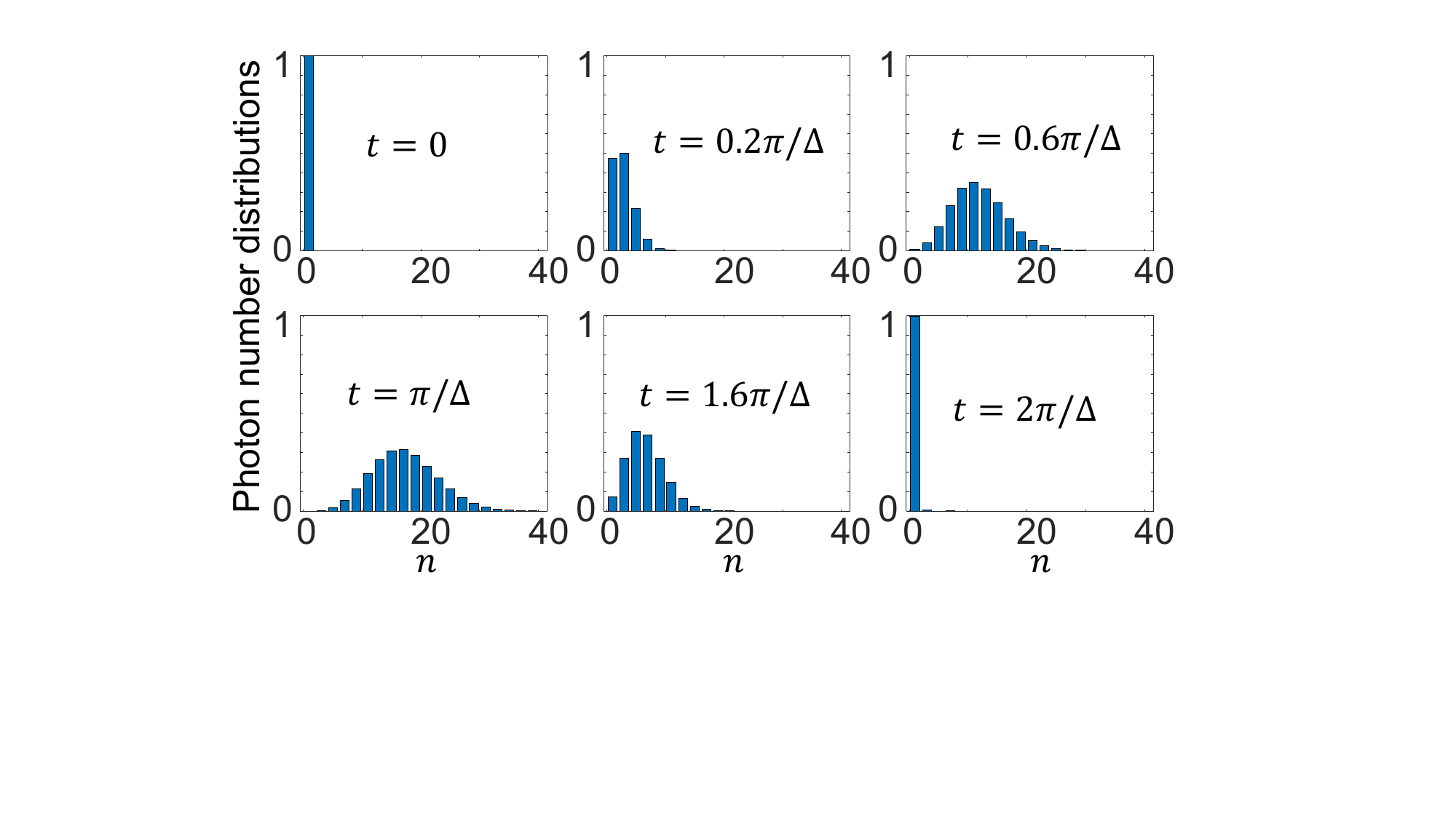}}
 	\caption{Instantaneous photon number distribution of the cavity mode $a$ in a finite-time evolution governed by $H$ with the initial state $|0,\mathcal{C}_{+}^{\beta}\rangle$, exploring the round trip of a photon number wave packet and collapse revivals. We choose parameters $\beta=2$, $\lambda=\Delta$, $K=10\Delta$, and $\delta=0$.
 	}
 	\label{fig2}
 \end{figure}
 
The effective Hamiltonian in Eq.~(\ref{eq3}) can be applied to study various physical phenomena.
Also, it is easy to achieve some generalizations of the QRM \cite{FornDiza2019,FriskKockum2019} by adding additional control fields.
The simplest application of
our approach is to enhance the coupling in a weak-coupling ($g\ll\Delta$) 
to the USC ($g\gtrsim 0.1\Delta$). 
The USC regime has a typical dynamical feature,
which is called ``collapses
and revivals'' \cite{Casanova2010PRL}. It describes 
the appearance of photon-number wave packets that bounce
back and forth along a defined parity chain, yielding collapses
and revivals of the initial population \cite{Casanova2010PRL}.
The parity chain is defined by the parity operator $\Pi=-\left(-1\right)^{a^{\dag}a}\sigma_{z}$ with
$\Pi|p\rangle=p|p\rangle$ and $p=\pm 1$.
For the initial state $|\psi(0)\rangle=|0,\mathcal{C}_{+}^{\beta}\rangle$ (corresponding to $p=+1$), {in the coupling regime with $g/\Delta\geq 1$ and $\tilde{\delta}=0$}, the coherent evolution of the system results in
\begin{align}
  |\psi(t)\rangle=\exp\left(\frac{i}{\Delta} g^{2}t\right)\exp{\left[-\frac{i}{\Delta^{2}}\sin(\Delta t)\right]}|\gamma(t),\mathcal{C}_{+}^{\beta}\rangle,
\end{align}
where $\gamma(t)=(g/\Delta)\left[\exp{(-i\Delta t)}-1\right]$ is the coherent state amplitude.
The revival probability of the initial state is $P_{+0}(t)=|\langle\psi(0)|\psi(t)\rangle|^{2}=\exp\left[-|\gamma(t)|^{2}\right]$,
which exhibits periodic collapses and full revivals as shown in Fig.~\ref{fig2}.
This demonstrates that we can effectively achieve the USC using the cat-state qubit.

 \begin{figure}[b]
	\centering
	\scalebox{0.35}{\includegraphics{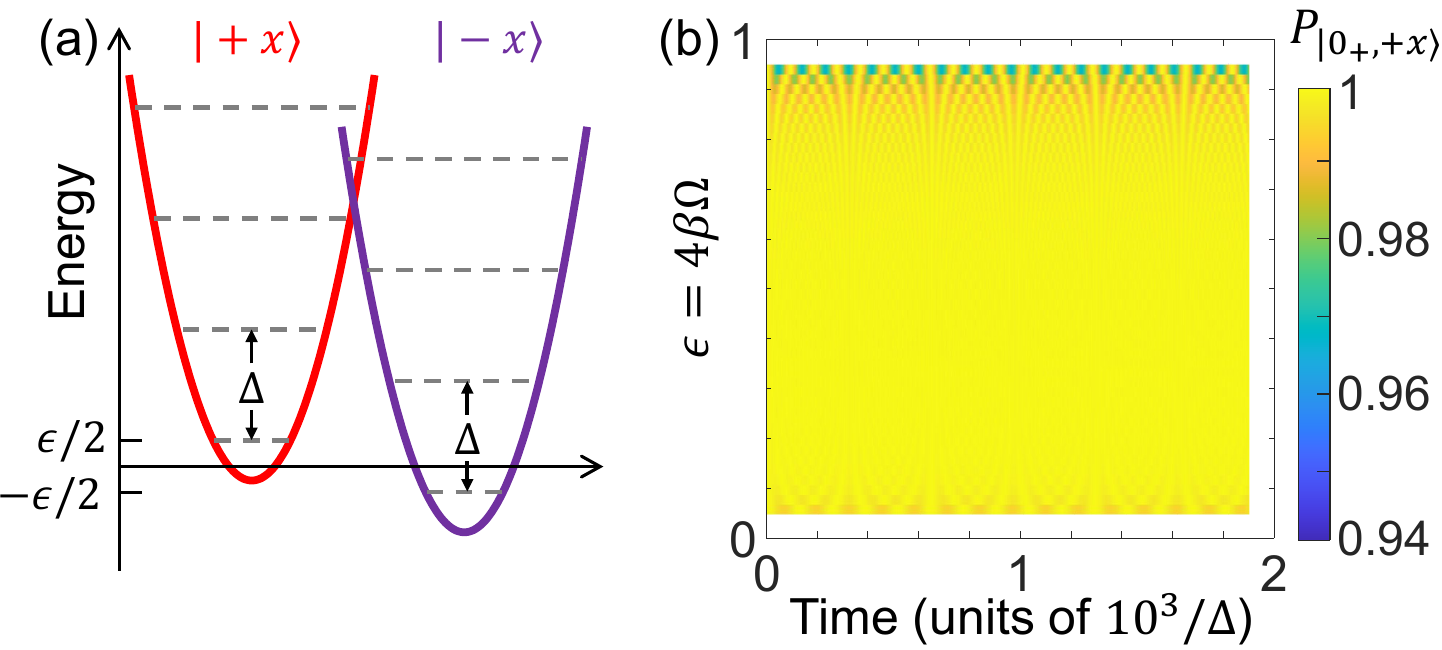}}
	\caption{(a) Schematic effective potentials of the asymmetric QRM for nonzero $\epsilon$, corresponding to a broken $\mathbbm{Z}_{2}$ symmetry. 
		(b) Time evolution governed by $H$ of the initial state $|0_{+},+x\rangle$ for $\epsilon\in[0.5,0.95]$. We choose $\beta=\sqrt{2}$ and $K=300\Delta$ for the cat-state qubit. Other parameters are $\lambda=0.5\Delta$ and $\tilde{\delta}=0.1\Delta$.
	}
	\label{fig3}
\end{figure}

\textit{Hidden symmetry and tunneling dynamics in the asymmetric QRM.}---
Assuming $|\beta|\geq2$ and $\lambda$ are real for simplicity, the asymmetric QRM can be obtained by applying a linear driving $H_{a}=\Omega(b+b^{\dag})$, with $\Omega\ll E_{\rm gap}$, onto the KNR, resulting in
\begin{align}
	H_{\rm AR}\simeq\Delta a^{\dag}a+\frac{\tilde{\delta}}{2}\sigma_{z}+\frac{\epsilon}{2}\sigma_{x}+g(a^{\dag}+a)\sigma_{x},
\end{align}
where $\epsilon\sigma_{x}/2$ is the additional asymmetric qubit bias term
with $\epsilon=4\beta\Omega$ and $\sigma_{x}=\sigma_{+}+\sigma_{-}$.
This additional $\sigma_{x}$ term breaks the $\mathbb{Z}_{2}$ symmetry in the standard QRM.
Level crossings appear in the
spectrum of the asymmetric QRM only if $\epsilon=n\Delta$ ($n=1,2,3,\ldots$) \cite{Wakayama2017Jpa,Ashhab2020Pra,Li2021Pra}.
These level crossings are expected to be associated with some hidden symmetry of the model \cite{Wakayama2017Jpa,Ashhab2020Pra}.
The origin of this hidden symmetry is established by finding the operators which
commute with the asymmetric QRM Hamiltonian at these special values.
Such a symmetry is obviously sensitive to deviations in the qubit bias $\sigma_{x}$ term.
In our protocol, the dominant error Hamiltonian in Eq.~(\ref{eq5}) does not contain 
off-diagonal elements (i.e., the $\sigma_{x}$ term) because the cat-state qubit preserves the
noise bias.

 \begin{figure}
	\centering
	\scalebox{0.34}{\includegraphics{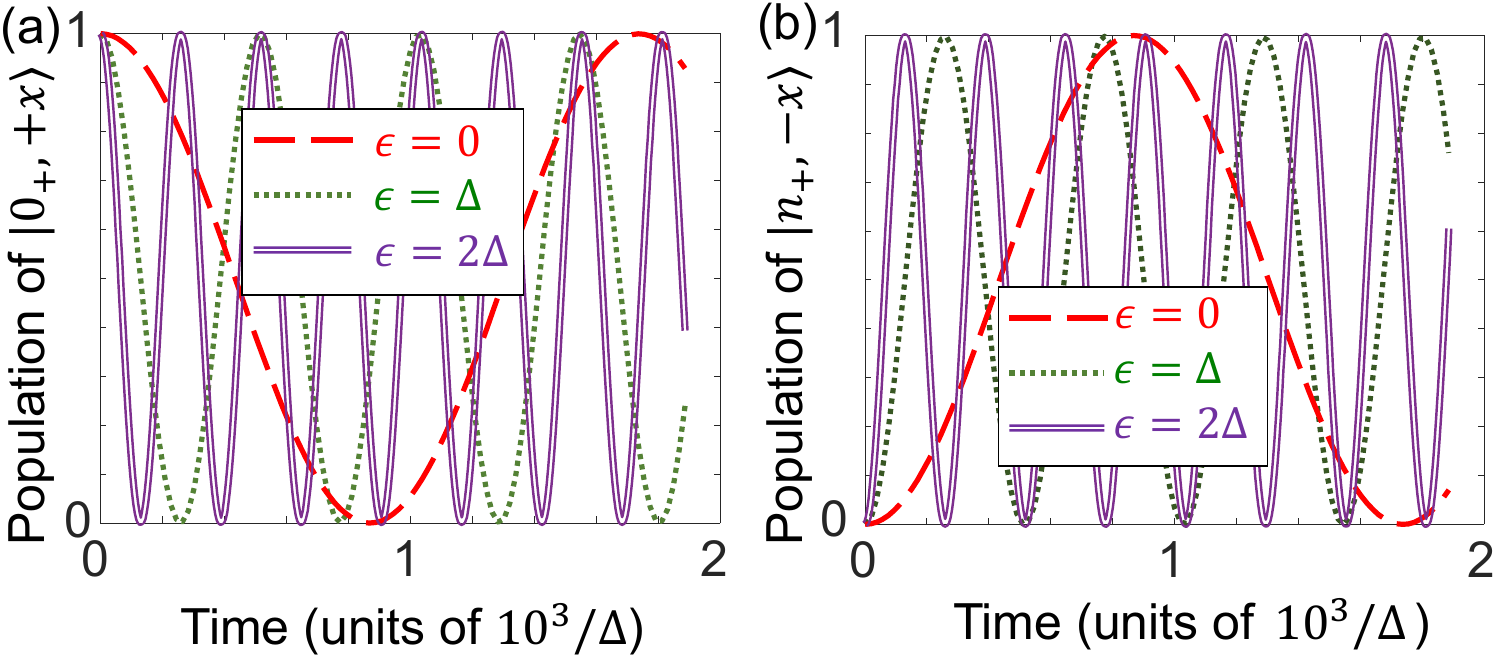}}
	\caption{Populations of (a) the initial state $|0_{+},+x\rangle$ and (b) the target states $|n_{-},-x\rangle$.
		For $\epsilon=n\Delta$, the tunneling process is reduced to a two level
		transition problem, resulting in a Rabi oscillation $|0_{+},+x\rangle\leftrightarrow|n_{-},-x\rangle$.
	    Parameters are the same as those in Fig.~\ref{fig3}(b).
	}
	\label{fig4}
\end{figure}

The existence of level crossings is independent of the value of $\tilde{\delta}$ 
when $\tilde{\delta}\neq 0$.
The term $H_{t}=\tilde{\delta}\sigma_{z}/2$, leading to transitions between
the eigenstates $|\pm x\rangle$ of $\sigma_{x}$, can be regarded as a tunneling term.
Removing the tunneling term $H_{t}$ from $H_{\rm AR}$, the rest of the Hamiltonian 
can be analytically solved with eigenstates
\begin{align}\label{eq6}
  |n_{\pm},\pm x\rangle\simeq D(\pm\alpha)|n\rangle\otimes \frac{1}{\sqrt{2}}\left(|\mathcal{C}_{+}^{\beta}\rangle\pm|\mathcal{C}_{-}^{\beta}\rangle\right),
\end{align}
where 
$D(\pm\alpha)=\exp{\left[\pm \alpha(a-a^{\dag})\right]}$ are displaced operators with amplitude $\alpha=g/\Delta\simeq\lambda\beta/\Delta$, 
and $|n\rangle$ are the Fock states. The corresponding 
eigenvalues are $E_{n}^{\pm}=n\Delta-g^{2}/\Delta\pm\epsilon/2$.

Equation.~(\ref{eq6}) shows the eigenstates of two displaced harmonic oscillators 
with displacing directions
determined by the two eigenvalues of $\sigma_{x}$.
The asymmetric qubit bias term $\epsilon \sigma_{x}/2$
lifts the degeneracy and leads to asymmetry
in the oscillator potentials \cite{Duan2015Epl} as shown in Fig.~\ref{fig3}(a). 
Thus, the levels $|m_{+},+x\rangle$ and $|(m+n)_{-},-x\rangle$
become degenerate when $\epsilon=n\Delta$. 
The tunneling process 
can be reduced to an analytically solvable two-level
resonant transition problem \cite{Irish2005Prb}.
The transition efficiency is determined by
the tunneling matrix elements $\langle m_{+}|(m+n)_{-}\rangle \tilde{\delta}/2$.
However, when $\epsilon$ is a non-integer multiple of $\Delta$, e.g., $\epsilon/\Delta\in[0.05,0.95]$,
the transition become off-resonance. 
Therefore, when $\epsilon/\Delta\in[0.05,0.95]$,
the system mostly remains in its initial state for a long time [see Fig.~\ref{fig3}(b)], indicating the tunneling probability decreases.
For $m=0$, a complete population transfer from $|0_{+},+x\rangle$ to $|n_{-},-x\rangle$
occurs when $\epsilon=n\Delta$,
indicating that the tunneling oscillation takes place (see Fig.~\ref{fig4}).

\textit{Pair-cat code.}---Using the bias term $\epsilon\sigma_{x}/2$ for control, and assuming $\tilde{\delta}=0$,
the lowest two eigenstates in Eq.~(\ref{eq6}) become degenerate. 
Their orthogonal basis 
\begin{align}
	|\mu_{\pm}\rangle=\frac{1}{\sqrt{2}}\left(|\alpha,+x\rangle\pm|-\alpha,-x\rangle\right),
\end{align}
are two-mode cat (or, pair-cat) states, which
can form a new computational subspace with a code projection
\begin{align}
  P_{\mu}=|\mu_{+}\rangle\langle\mu_{+}|+|\mu_{-}\rangle\langle\mu_{-}|.
\end{align}
Similar to the single-cat qubits \cite{Puri2017,PuriPrx2019,Cai2021}, our pair-cat qubit
can also, even more effectively, preserve the
noise bias.
Focusing on error operators $b$ and $b^{\dag}b$, the 
bias-preserving parameters
\begin{align}\label{eq11}
	&\langle\mu_{+}|b^{\dag}b|\mu_{+}\rangle-\langle\mu_{-}|b^{\dag}b|\mu_{-}\rangle=|\beta|^{2} e^{-2|\alpha|^2}\left(A^{-2}-A^{2}\right),\cr
	&\langle\mu_{+}|b|\mu_{-}\rangle-\langle\mu_{-}|b|\mu_{+}\rangle=\beta e^{-2|\alpha|^2}\left(A-A^{-1}\right),
\end{align}
are exponentially smaller than those of the cat-state qubit:
\begin{align}\label{eq12}
	&\langle \mathcal{C}_{+}^{\beta}|b^{\dag}b|\mathcal{C}_{+}^{\beta}\rangle-\langle\mathcal{C}_{-}^{\beta}|b^{\dag}b|\mathcal{C}_{-}^{\beta}\rangle=|\beta|^{2}\left(A^{-2}-A^{2}\right),\cr
	&\langle\mathcal{C}_{+}^{\beta}|b|\mathcal{C}_{-}^{\beta}\rangle-\langle\mathcal{C}_{-}^{\beta}|b|\mathcal{C}_{+}^{\beta}\rangle=\beta \left(A-A^{-1}\right).
\end{align}
For operators $a$ and $a^{\dag}a$, we have $\langle\mu_{+}|a|\mu_{-}\rangle=\langle\mu_{-}|a|\mu_{+}\rangle$ 
and  $\langle\mu_{+}|a^{\dag}a|\mu_{+}\rangle=\langle\mu_{-}|a^{\dag}a|\mu_{-}\rangle$. 
These indicate that the pair-cat code can satisfy the Knill-Laflamme condition \cite{Bennett1996Pra,Knill1997Pra}
better than the single-cat code regarding single-photon-loss error.
Moreover, Eq.~(\ref{eq11}) demonstrates that 
a projection of the error Hamiltonian $H_{\rm err}$ onto the pair-cat subspace using $P_{\mu}$ also
results in a unit matrix for large $\alpha$ and $\beta$. 
Therefore, the simulated QRM can be a great candidate for realizing fault-tolerant codes tailored to
biased-noise qubits.

\begin{figure}
	\centering
	\scalebox{0.34}{\includegraphics{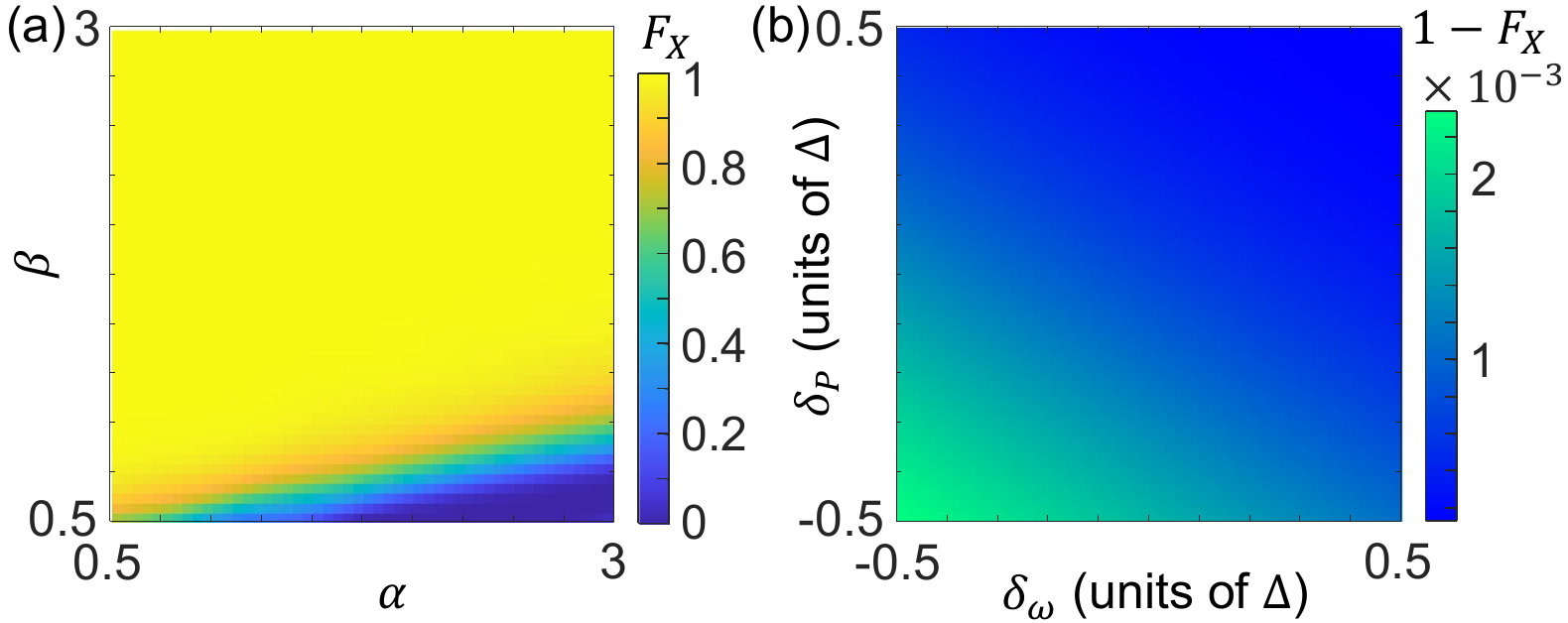}}
	\caption{(a) Average fidelity $F_{X}$ of the pair-cat-code Paul-$X$ gate versus $\alpha$ and $\beta$ when $K=10\Delta$ and $\delta=0$. (b) Average infidelity ($1-F_{X}$) of the Paul-$X$ gate in the presence of the error Hamiltonian $H_{\rm err}$. We choose $\alpha=\beta$ to be real and  $\lambda=\Delta=0.1K$ to satisfy $\lambda,\Delta\ll E_{\rm gap}$.
	}
	\label{fig5}
\end{figure}

\textit{Paul-$X$ gate.}---Noting that $a$ and $b$ are both uncorrectable errors, 
we can apply the control term $\epsilon\sigma_{x}/2$ to the system to create an $X$ gate. 
In the limit $\epsilon\ll E_{\rm gap}$, 
this additional drive lifts the degeneracy between the states $|\alpha,+x\rangle$ and $|-\alpha,-x\rangle$ and leads to oscillations with an effective Rabi frequency 
$\epsilon=4\beta\Omega$
between the states $|\mu_{\pm}\rangle$. Choosing an evolution time $t_{\rm gate}=\pi/\epsilon$,
the evolution operator of the system becomes
\begin{align}
	U_{X}=|\mu_{+}\rangle\langle\mu_{-}|+|\mu_{-}\rangle\langle\mu_{+}|,
\end{align}
i.e., the Paul-$X$ gate.
The average fidelity of the Paul-$X$ gate over all possible
initial states can be defined by \cite{Zanardi2004Pra}
\begin{align}
	F_{X}=\frac{{\rm Tr}(MM^{\dag})+|{\rm Tr}(M)|^2}{d^{2}+d}, 
\end{align}
where $M=\mathcal{P}_{c}U_{X}^{\dag}U(t_{\rm gate})\mathcal{P}_{c}$, with $\mathcal{P}_{c}$ ($d$)
being the projector (dimension) of the computational subspace $\mathcal{C}_{\mu}=\left\{|\mu_{\pm}\rangle\right\}$. 
Here, $U(t_{\rm gate})$ is the actual evolution operator calculated
for the Hamiltonian $H$.
The gate fidelities calculated for different $\alpha$ and $\beta$ are
shown in Fig.~\ref{fig5}(a). 
As $\alpha$ and $\beta$ increase, the gate fidelity increases when choosing a fixed $\Delta$.
Noting that a larger $\alpha$ corresponds to a 
larger $\lambda$ for fixed $\beta$, this can lead to infidelity 
because the condition $\lambda\ll E_{\rm gap}$ cannot be well satisfied. 
This can cause population leakage out of the computational subspace and reduce the gate fidelity.

\begin{figure}
	\centering
	\scalebox{0.28}{\includegraphics{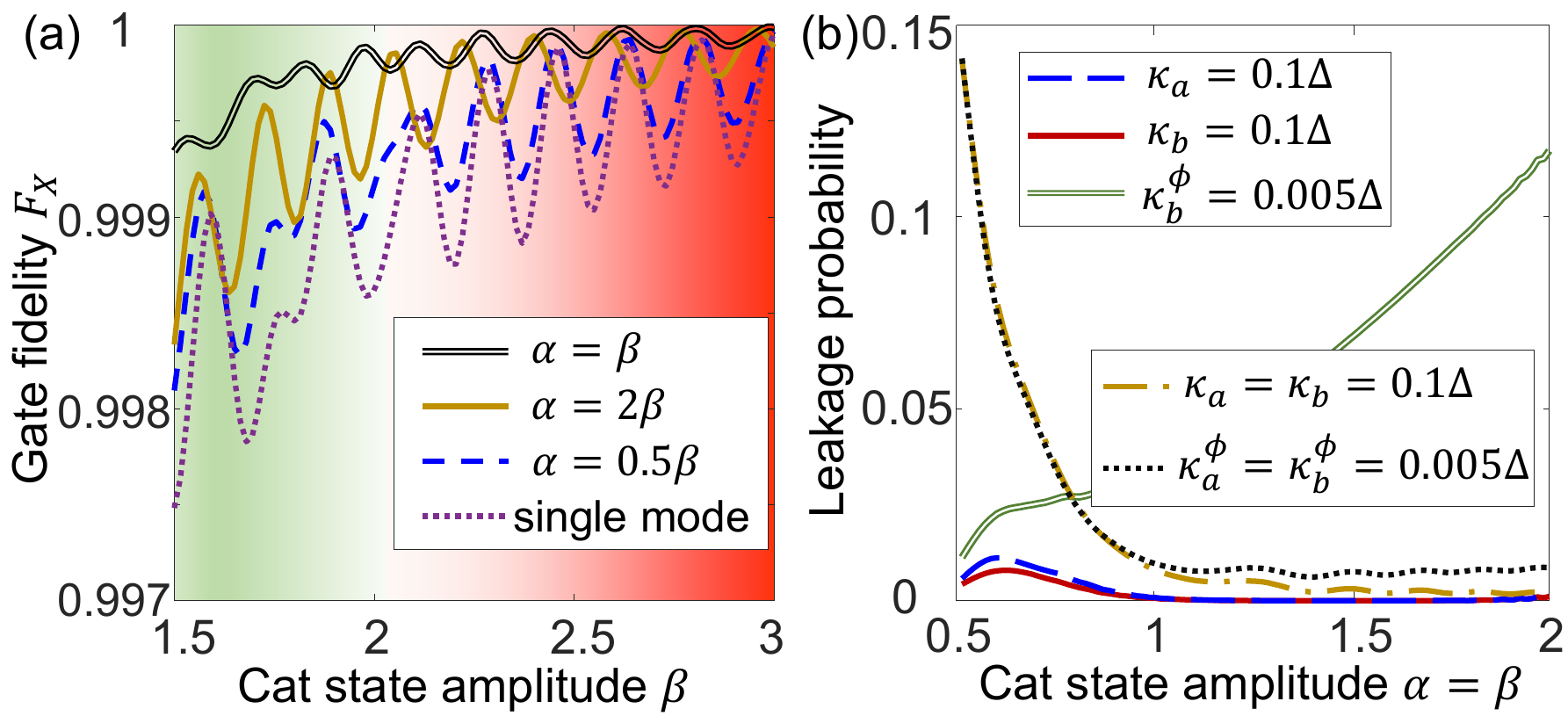}}
	\caption{(a) Comparison of $F_{X}$ between the pair-cat and the single-cat codes. 
		(b) Probabilities of excitations out of the $\mathcal{C}$ and $\mathcal{C}_{\mu}$ subspaces, given respectively by $\left[1-\langle \mathcal{C}_{+}^{\beta}|\rho|\mathcal{C}_{+}^{\beta}\rangle-\langle \mathcal{C}_{-}^{\beta}|\rho|\mathcal{C}_{-}^{\beta}\rangle\right]$ 
		and $\left[1-\langle \mu_{+}|\rho|\mu_{+}\rangle-\langle \mu_{-}|\rho|\mu_{-}\rangle\right]$ in the presence of decoherence. Other parameters are $\lambda=\Delta=0.1K$ and $\tilde{\delta}=0.01\Delta$. For the pair-cat code, we choose $\delta=0$. 
	}
	\label{fig6}
\end{figure}

The projection of $H_{\rm err}$ onto the pair-cat subspace
also results in a nearly unit matrix for large $\beta$, indicating the robustness of the pair-cat 
Paul-$X$ gate against parameter imperfections in $P$ and $\delta$.
As shown in Fig.~\ref{fig5}(b), the error Hamiltonian $H_{\rm err}$ only causes
$\lesssim0.2\%$ infidelity to the Paul-$X$ gate, even when the deviations $\delta_{P}$ and $\delta_{\omega}$ reach $\pm0.5\Delta$. 
Moreover, because the bias-preserving parameters in Eq.~(\ref{eq11}) are exponentially smaller than those
in Eq.~(\ref{eq12}), the influence of $H_{\rm err}$ is also exponentially suppressed in 
the pair-cat protocol.
A comparison [see Fig.~\ref{fig6}(a)] between our pair-cat protocol 
and the single-cat one \cite{Grimm2020}
indicates that
our protocol can more efficiently suppress parameter deviations in the parametric drive.  
Specifically, choosing $\alpha=\beta$, the gate fidelity mostly 
remains in $F_{X}\geq 99.95\%$ [black-hollow curve in Fig.~\ref{fig6}(a)].

\textit{Decoherence.}---In the presence of single-photon losses and dephasing, 
the system dynamics
is described by the Lindblad master equation
\begin{align}
	\dot{\rho}=-i[H,\rho]+\sum_{j=a,b}\kappa_{j}\mathcal{D}[j]\rho+\kappa_{j}^{\phi}\mathcal{D}[j^{\dag}j]\rho,
\end{align}
where $\mathcal{D}[o]\rho=o\rho o^{\dag}-\left(o^{\dag}o\rho+\rho o^{\dag}o\right)/2$ is the standard Lindblad
superoperator and $\kappa_{j}$ ($\kappa_{j}^{\phi}$) is the single-photon loss (dephasing) rate
of the cavity mode $j$ with $j=a,b$.
Assuming $\kappa_{j},\kappa_{j}^{\phi}\ll E_{\rm gap}$, the dynamics of the cat-state qubit is well
confined to the subspace $\mathcal{C}$ because a stochastic
jump does not cause leakage to the
excited eigenstates for large $\beta$ \cite{PuriPrx2019,Puri2020,Chen2022Prappl}.

After a projection of the system onto the cat-state subspace $\mathcal{C}$, the effective master equation becomes (see details in the Supplemental Material \cite{SM})
\begin{align}
	\dot{\rho}\approx &-i[H_{R},\rho]+\kappa\mathcal{D}[a]\rho+\kappa_{a}^{\phi}\mathcal{D}[a^{\dag}a]\rho\cr
	&+\kappa_{a}|\beta|^2\mathcal{D}\left[\frac{A+A^{-1}}{2}\sigma_{x}+\frac{A-A^{-1}}{2}\sigma_{y}\right]\rho\cr
	&+\kappa_{b}^{\phi}|\beta|^{4}\mathcal{D}\left[\frac{A^{2}+A^{-2}}{2}\mathbbm{1}_{\beta}-\frac{A^{2}-A^{-2}}{2}\sigma_{z}\right]\rho.
\end{align}
For large $\beta$, the $\sigma_{y}$ and $\sigma_{z}$ terms are exponentially suppressed, leaving only the bit-flipping error $\sigma_{x}$.
As shown in Fig.~\ref{fig6}(b), when considering only single-photon losses, the probability to go out of the cat-state subspace
is negligible (see the blue-dashed and red-solid curves).
However, because ${b^{\dag}b|\pm\beta\rangle=\beta^{2}|\pm\beta\rangle\pm\beta D(\pm\beta)|1\rangle}$,
the leakage probability becomes proportional to
$\left(\kappa_{b}^{\phi}\beta/E_{\rm gap}\right)^{2}$ for pure dephasing (the green-hollow curve) \cite{PuriPrx2019,SM}. To suppress such a leakage, 
the dephasing rate should be as small as possible for the simulation protocol.

For the pair-cat code, according to Eqs.~(\ref{eq11}) and (\ref{eq12}), one can calculate that 
single-photon losses cannot induce leakage out of the computational subspace $\mathcal{C}_{\mu}$ when $\alpha=\beta\geq \sqrt{2}$. 
This is demonstrated with the brown-dashed-dotted curve in Fig.~\ref{fig5}(b).
Single-photon losses only induce bit-flip error. 
Similar to the case of a single-cat qubit, the pair-cat qubit is also robust against phase-flip error,
as demonstrated in Eqs.~(\ref{eq11},\ref{eq12}). 
Though, the dephasing rates should be small to 
suppress the leakage probability. For instance, the leakage probability is about $0.5\%$ when $\kappa_{a}^{\phi}=\kappa_{b}^{\phi}=0.005\Delta$.
Enlarging the amplitudes $\alpha$ and $\beta$ can suppress the leakage probability because
${a^{\dag}a|\pm\alpha\rangle=\alpha^{2}|\pm\alpha\rangle\pm\alpha D(\pm\alpha)|1\rangle}$ and 
${b^{\dag}b|\pm\beta\rangle=\beta^{2}|\pm\beta\rangle\pm\beta D(\pm\beta)|1\rangle}$ (see details in the Supplemental Material \cite{SM}).
However, this increases the experimental difficulty in realizing the protocol.

\textit{Conclusions.}---We have investigated how to amplify the coupling between
a parametrically-driven KNR (corresponding to a cat-state qubit) and cavity
to effectively reach the USC.
The bias-preserving character of the cat-state qubit makes the simulation protocol
robust against the frequency mismatch and the amplitude mismatch of the parametric drive.
Thus, a precise effective Hamiltonian can be obtained for exploring USC-induced quantum phenomena and applications, 
such as collapse and revivals of quantum states, pair-cat-code computation, as well as hidden symmetry and tunneling dynamics. Our numerical simulations show that this protocol can simulate the
USC dynamics with high fidelity in the presence of parameter imperfections. 
We have applied the model for implementing a pair-cat code, which is a promising error-correction code because 
it meets the Knill-Laflamme condition better and preserves the
noise bias stronger than a single-cat code. 
{This allows to reach the same level of protection of single-cat codes with a 
lower average photon number per mode. We can predict that further increasing the 
number of modes can further reduce the average photon number per mode to reach the same level of protection \cite{Albert2019}. 
However, this may make the system too complicated to realize.}
In summery, our results open a path toward error-tolerant simulations of ultrastrong light matter couplings, as well as promising applications of error-correction qubits \cite{Girvin2023Spln}.

\begin{acknowledgments}

Y.-H.C. is supported by the National Natural Science Foundation of China
under Grant No. 12304390.
Y.X. is supported
by the National Natural Science Foundation of China
under Grant No. 11575045, the Natural Science Funds
for Distinguished Young Scholar of Fujian Province under
Grant 2020J06011 and Project from Fuzhou University
under Grant JG202001-2.
F.N. is supported in part by: 
Nippon Telegraph and Telephone Corporation (NTT) Research, 
the Japan Science and Technology Agency (JST) 
[via the Quantum Leap Flagship Program (Q-LEAP), and the Moonshot R\&D Grant Number JPMJMS2061], 
the Asian Office of Aerospace Research and Development (AOARD) (via Grant No. FA2386-20-1-4069), 
and the Office of Naval Research (ONR) (via Grant No. N62909-23-1-2074).

\end{acknowledgments}

%\end{CJK*}
\bibliography{references}

\end{document}